\documentclass[12pt]{iopart}
\usepackage{iopams}  

\usepackage{graphicx}
\usepackage{dcolumn}
\usepackage{bm}

\usepackage{amssymb}

\begin{document}

\newcommand{\cm}{cm$^{-1}$} \newcommand{\A}{\AA$^{-1}$} \newcommand{\Q}{\mathbf{Q}} 

\title[Macroscopically entangled protons] {Evidence of macroscopically entangled protons in a mixed isotope crystal of KH$_{p}$D$_{1-p}$CO$_3$ }

\author{Fran\c{c}ois Fillaux}
\footnote[3]{To whom correspondence should be addressed:\\ fillaux@glvt-cnrs.fr\\ Tel: +33149781283\\ Fax: +33149781218 \\ (fillaux@glvt-cnrs.fr; http://www.ladir.cnrs.fr/pagefillaux\_eng.htm)}
\address{LADIR-CNRS, UMR 7075,
Universit\'{e} P. et M. Curie, 2 rue Henry Dunant, 94320 Thiais, France}
\author{Alain Cousson}
\address{Laboratoire L\'{e}on Brillouin (CEA-CNRS), C.E. Saclay, 91191
Gif-sur-Yvette, cedex, France}
\author{Matthias J Gutmann}
\address{ISIS Facility, Rutherford Appleton Laboratory, Chilton, Didcot,
OX11 0QX, UK}
\date{\today}

\begin{abstract}
We examine whether protons and deuterons in the crystal of KH$_{0.76}$D$_{0.24}$CO$_3$ at 300 K are particles or matter waves. The neutron scattering function measured over a broad range of reciprocal space reveals the enhanced diffraction pattern anticipated for antisymmetrized macroscopic states for protons (fermions). These features exclude a statistical distribution of protons and deuterons. Raman spectra are consistent with a mixture of KHCO$_3$ and KDCO$_3$ sublattices whose isomorphous structures are independent of the isotope content. We propose a theoretical framework for decoherence-free proton and deuteron states. 
\end{abstract}


\maketitle

\section{Introduction}
A defect-free crystal is a macroscopic quantum system with spatial periodicity. In principle, only nonlocal observables are probed by plane waves, eg. with diffraction or spectroscopy techniques. This is attested by the marked localization in reciprocal space of Bragg peaks as well as discrete phonon states. However, since the seminal work of Slater \cite{Slater}, the nonlocal nature of protons is commonly overlooked in crystals containing bistable O--H$\cdots$O hydrogen bonds, when the coexistence of two configurations, say ``$L$'' for $\mathrm{O1-H}\cdots \mathrm{O2}$ and ``$R$'' for $\mathrm{O1}\cdots \mathrm{H-O2}$, is conceived of as classical disorder. This leads to a conflict of interpretation consisting in that a statistical distribution of protons should destroy the translation periodicity of the crystal, whereas infrared and Raman spectra evidence extended states at the Brillouin-zone center (BZC) obeying the symmetry-related selection rules of the crystal. This conflict has been tolerated because the nonlocal nature of protons cannot be evidenced by x-ray diffraction. Even with neutrons, a Bragg peak analysis is not conclusive, for statics and dynamics cannot be unravelled effectively from the unit-cell structure factor \cite{Nelmes2}. 

The same conflict of interpretation arises for crystals containing hydrogen bonded centrosymmetric dimers, such as carboxylic acids \cite{WSF1,ST}, or potassium hydrogen carbonate (or bicarbonate, KHCO$_3$) \cite{TTO1,TTO2,BHT,EGS}. The thermal interconversion between tautomeric configurations, $LL$ and $RR$, is supposed to be correlated with a reorganization of the single and double CO bonds, along a reaction path across a multidimensional potential. This is commonly rationalized with rigid proton-pairs in an asymmetric double-well coupled to an incoherent thermal bath mimicking a liquid-like environment. It is supposed that ``phonon-induced incoherent tunnelling'' should prevail at low temperatures, whereas stochastic jumps should take the lead at elevated temperatures. By contrast, vibrational spectra are consistent with nonlocal observables represented by normal coordinates. There is no evidence of any symmetry breaking due to local disorder and the density-of-states remains discrete at any temperature \cite{LN,FLR,FRLL}. 

Nowadays, the local or nonlocal behaviour of bistable hydrogen bonds can be distinguished at advanced pulsed neutron sources, with a time-of-flight Laue diffractometer measuring a much broader range of reciprocal space than any reactor-based four-circle instrument. The scattering function measured at large momentum transfer values ($\Q = \mathbf{k}_i - \mathbf{k}_f$, where  $\mathbf{k}_i$ and $\mathbf{k}_f$ are the initial and final wave vectors, respectively) contains information on the scale of spacial correlations between $L$ and $R$ sites ($\sim 0.5$ \AA). Moreover, such an instrument is unique to observing diffuse scattering, because signals in addition to Bragg peaks are effectively measured across the whole range of reciprocal space for each neutron pulse. It is thus possible to determine whether or not coherent scattering is affected by the coexistence of $L$ and $R$ configurations at various temperatures. 

This has been exemplified with the crystal of KHCO$_3$ made of hydrogen bonded dimers, (HCO$_3^-)_2$ \cite{FCKeen,FCG2,FCG4,FCG5}. Below 150 K, only those proton sites ($L$), corresponding to a unique configuration for dimers ($LL$), are occupied. The neutron scattering function reveals enhanced Bragg peaks exclusively at those particular $\Q$-values corresponding to the reciprocal sublattice of protons. Furthermore, the enhanced peaks are superposed to a prominent continuum of incoherent scattering, cigar-like shaped along $\Q$-vectors perpendicular to dimer planes. By contrast, there is no evidence of similar enhanced features for the isomorphous derivative KDCO$_3$. To the best of our knowledge, the dramatic increase of both coherent and incoherent scattering intensities for the sublattice of protons, not observed for deuterons, cannot be explained with any conventional model based upon Slater's localized particles. Alternatively, a quantum theory leads to different macroscopic quantum correlations (entanglement) for indistinguishable protons (fermions) or deuterons (bosons). The enhanced features can be thus explained by the spin symmetry of the ``super-rigid'' fermion states for which phonons are forbidden. By contrast, deuteron states are represented by phonons with no spin-symmetry. 

Above 150 K, the less favored sites along the hydrogen bonds ($R$), at $\Delta_x \approx 0.6$ \AA\ from the $L$ sites, are progressively populated, while the space group symmetry and the unit cell parameters remain unchanged. At 300 K, the population ratio is $\approx 18(RR):82(LL)$. The enhanced features observed continuously from cryogenic to room temperatures show that the quantum correlations are temperature independent. The coexistence of $LL$ and $RR$ configurations is consistent with a quantum superposition of proton states perfectly insulated from decoherence by the other degrees of freedom \cite{FCG2,FCG4}. Then, scattering functions, vibrational spectra, and quasi-elastic neutron scattering (QENS) data, can be reinterpreted consistently with nonlocal observables and quantum interference arising from the superposition of macroscopic tunnelling states. 

Structural transitions of second order ($P2_1/a \longleftrightarrow C2/m$) occur at $T_{cH} \approx 318$ K (KHCO$_3$) or $T_{cD} \approx 353$ K (KDCO$_3$). Above $T_c$, dimers split into two centrosymmetric entities, symmetric with respect to the plane $(a,c)$. It is quite remarkable that the enhanced features of the scattering function, as well as proton dynamics and quantum interference, do not show any discontinuity at $T_c$ \cite{FCG4}. There is no evidence of classical disorder above $T_c$. It transpires that quantum correlations arise from the space periodicity of centrosymmetric dimers, irrespective of the space group symmetry. 

Here, we examine whether the enhanced features survive, or not, in a mixed isotope crystal of KH$_{p}$D$_{1-p}$CO$_3$. In this case, static or dynamical disorder are clearly differentiated and the existence of distinct crystallographic domains of KHCO$_3$ or KDCO$_3$ is excluded. A static distribution of mutually exclusive particles should destroy the quantum entanglement and the enhanced features should vanish. Alternatively, if isotopes merge into separable macroscopic states, such that quantum correlations across the sublattice of protons survive, the enhanced features should be observed, although with lower contrast compared to KHCO$_3$. To the best of our knowledge, such a nonlocal approach has never been examined in depth, in this context. 

In a preliminary report \cite{FCou2}, the structure of KH$_{0.76}$D$_{0.24}$CO$_3$ at 300 K was refined from Bragg peaks measured with the four-circle diffractometer 5C4 at the Orph\'{e}e reactor of the Laboratoire L\'{e}on Brillouin \cite{LLB}. There was no evidence of any symmetry breaking: the space group ($P2_1/a$) is identical to that of KHCO$_3$ or KDCO$_3$ at the same temperature (Fig. \ref{fig:1}). A structural model consistent with the proposed quantum theory was represented by four sublattices, H$_{LL}$, H$_{RR}$, D$_{LL}$, D$_{RR}$, with weights $p_Hq_{LLH}, p_Hq_{RRH}, p_Dq_{LLD}, p_Dq_{RRD}$, respectively: 
\begin{equation}\label{eq:1}
\begin{array}{l}
p_H [q_{LLH}\mathrm{H}_{LL} + q_{RRH} \mathrm{H}_{RR}] + p_D [q_{LLD}\mathrm{D}_{LL} + q_{RRD} \mathrm{D}_{RR}]. 
\end{array}\end{equation}
The bound coherent scattering lengths are $\bar{b}_H = -3.741$ fm or $\bar{b}_D = 6.671$ fm. The refined positional and thermal parameters ($R \approx 3\%$) were identical for protons and deuterons whose sites are indiscernible, to within experimental precision and thermal ellipsoids. The estimated concentrations were $p_H = 0.76$, $p_D = 0.24$, and the relative weights were such as $S = p_Hq_{LLH} + p_Hq_{RRH} + p_Dq_{LLD} + p_Dq_{RRD} = 0.99(4)$. There was no evidence of any significant decrease of Bragg peak intensities into Laue monotonic diffuse scattering due to a statistical distribution of isotopes. Alternatively, for a sublattice of paired up nonseparable protons and deuterons, the averaged scattering length should be: 
\begin{equation}\label{eq:2}
\bar{b} = 0.76 \bar{b}_H + 0.24 \bar{b}_D \approx -1.242 \ \mathrm{fm}, 
\end{equation}
The model is more artificial because $p$ is no longer a free parameter and it is less satisfactory ($R \approx 4\%$). However, it cannot be definitely excluded. Consequently, the question as to whether the sublattices of protons and deuterons are separable entities remains largely open. 

The conclusions of the present work are based upon the neutron scattering function of the mixed isotope crystal presented in Sec. \ref{sec:2}. The enhanced features show that the quantum correlations of the pure crystal are not destroyed by partial deuteration and Raman spectra are consistent with separable proton and deuteron states. In Sec. \ref{sec:3}, the quantum theory \cite{FCG4,FCG5} applied to the mixed crystal accounts for macroscopic proton and deuteron states, each of which being a superposition of $LL$ and $RR$ states. A detailed interpretation of the enhanced features is proposed. In Sec. \ref{sec:4}, we show that alternative explanations are inappropriate. 

\section{\label{sec:2}Experimental}

\subsection{\label{sec:2A}The neutron scattering function}

The mixed isotope crystal ($\approx 30$ mm$^3$) measured in the previous diffraction work \cite{FCou2} was cut from a twinning-free parent crystal ($\approx 20\times 15 \times 3$ mm$^3$) obtained from a water solution ($\approx 75 \%$ H$_2$O, $25\%$ D$_2$O). This parent crystal was loaded in the vacuum chamber of the time-of-flight Laue diffractometer SXD, at the ISIS pulsed neutron source \cite{SXD,SXD2}. The temperature was ($300 \pm 1$) K. We measured 8 orientations, each for $\approx 8$ hours, with $(b) \parallel (b^*)$ either parallel or perpendicular to the equatorial plane of the detector bank. The reduced data were concatenated in order to compute the scattering function (Fig. \ref{fig:2}). Previous studies of crystals with similar sizes have shown that multiple-scattering effects are not important. However, Bragg peaks show severe extinction. A standard analysis is consistent with the structure and isotope contents previously determined, but the modest precision ($R \approx 10\%$) shows that measurements conducted at different sources are complementary. 

\begin{figure}\begin{center}
\includegraphics[scale=0.5]{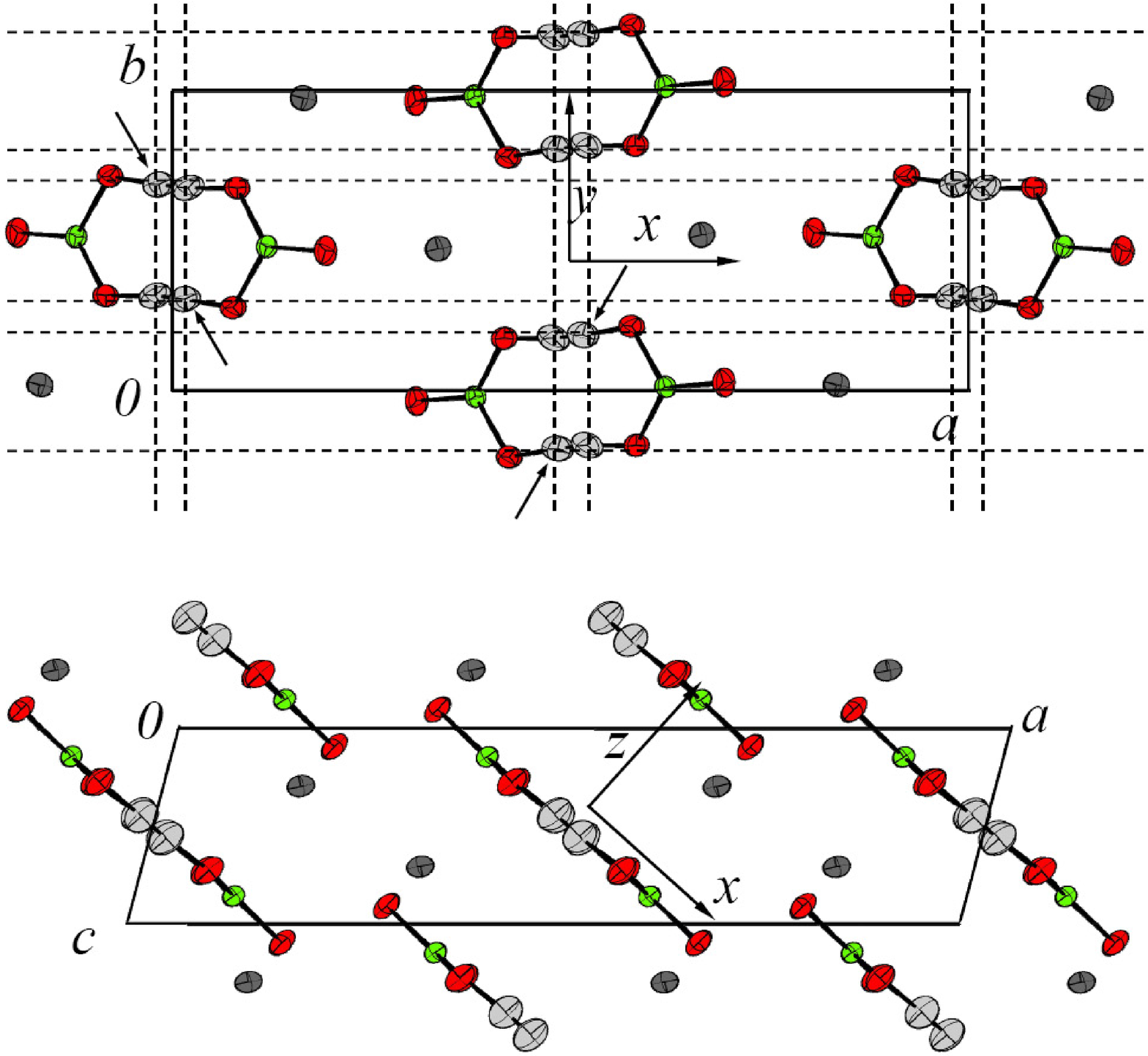}
\caption{\label{fig:1} (Color on line) Schematic view of the crystalline structure of KH$_{0.76}$D$_{0.24}$CO$_3$ at 300 K. Top: projection onto $(a,b)$ planes. Bottom: projection onto $(a,c)$ planes. The solid lines represent the unit cell. Proton and deuteron sites are indiscernible within thermal ellipsoids. The dashed lines emphasize the sublattice of protons/deuterons. The arrows point to the sites populated at low temperature. }
\end{center}\end{figure}

\begin{figure}\begin{center}
\includegraphics[angle=0.,scale=0.40]{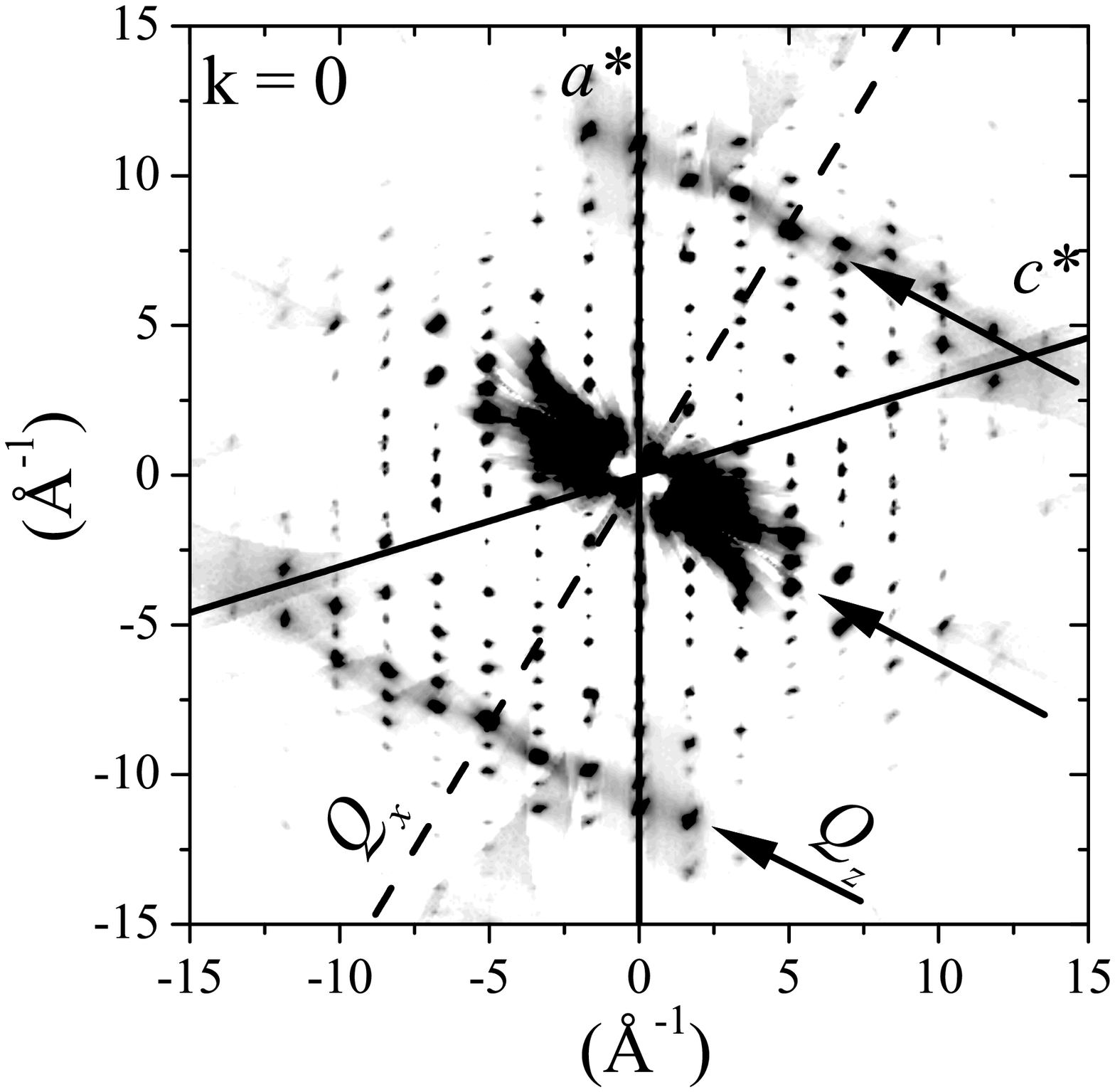}
\includegraphics[angle=0.,scale=0.40]{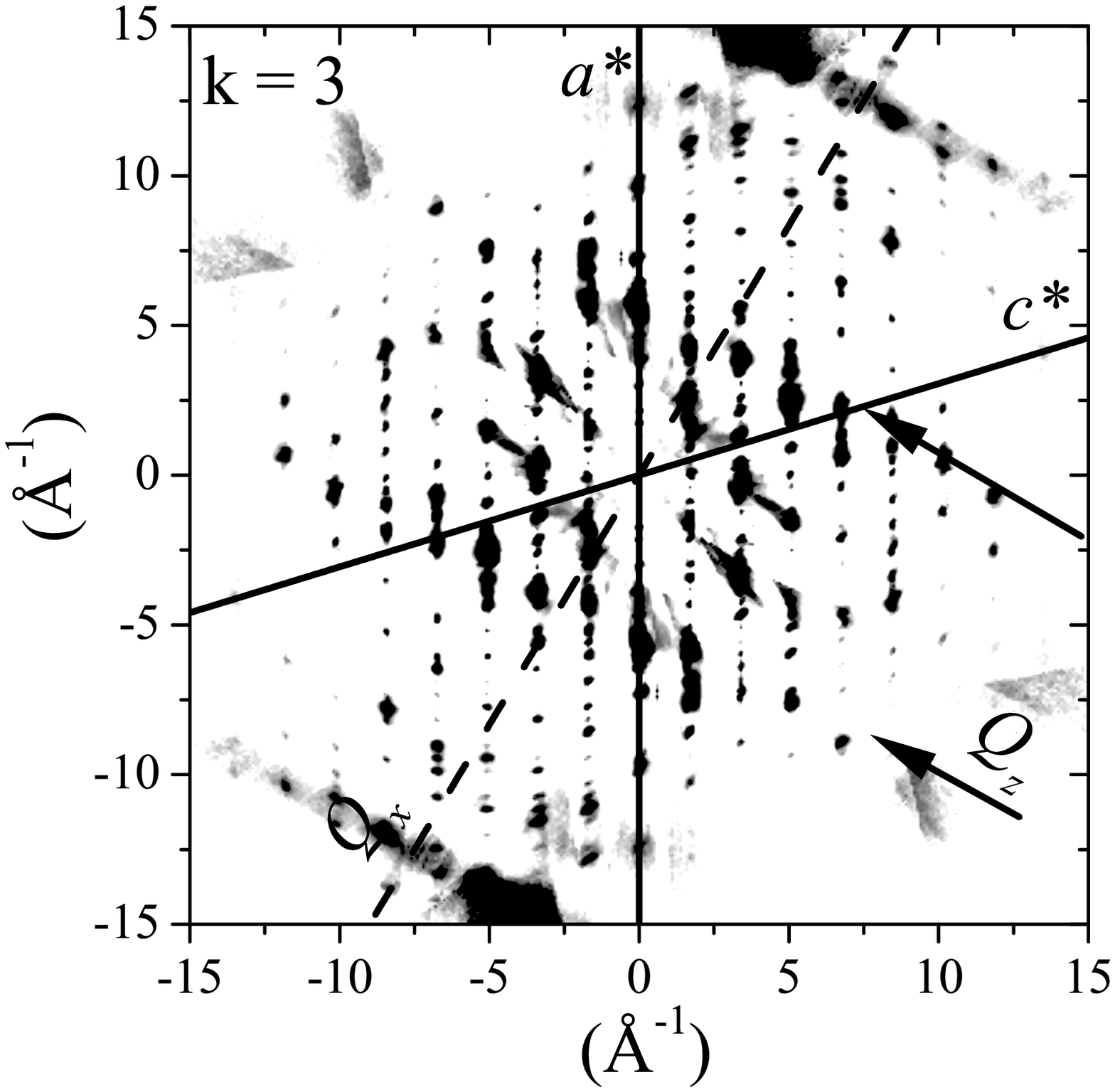}\\
\includegraphics[angle=0.,scale=0.40]{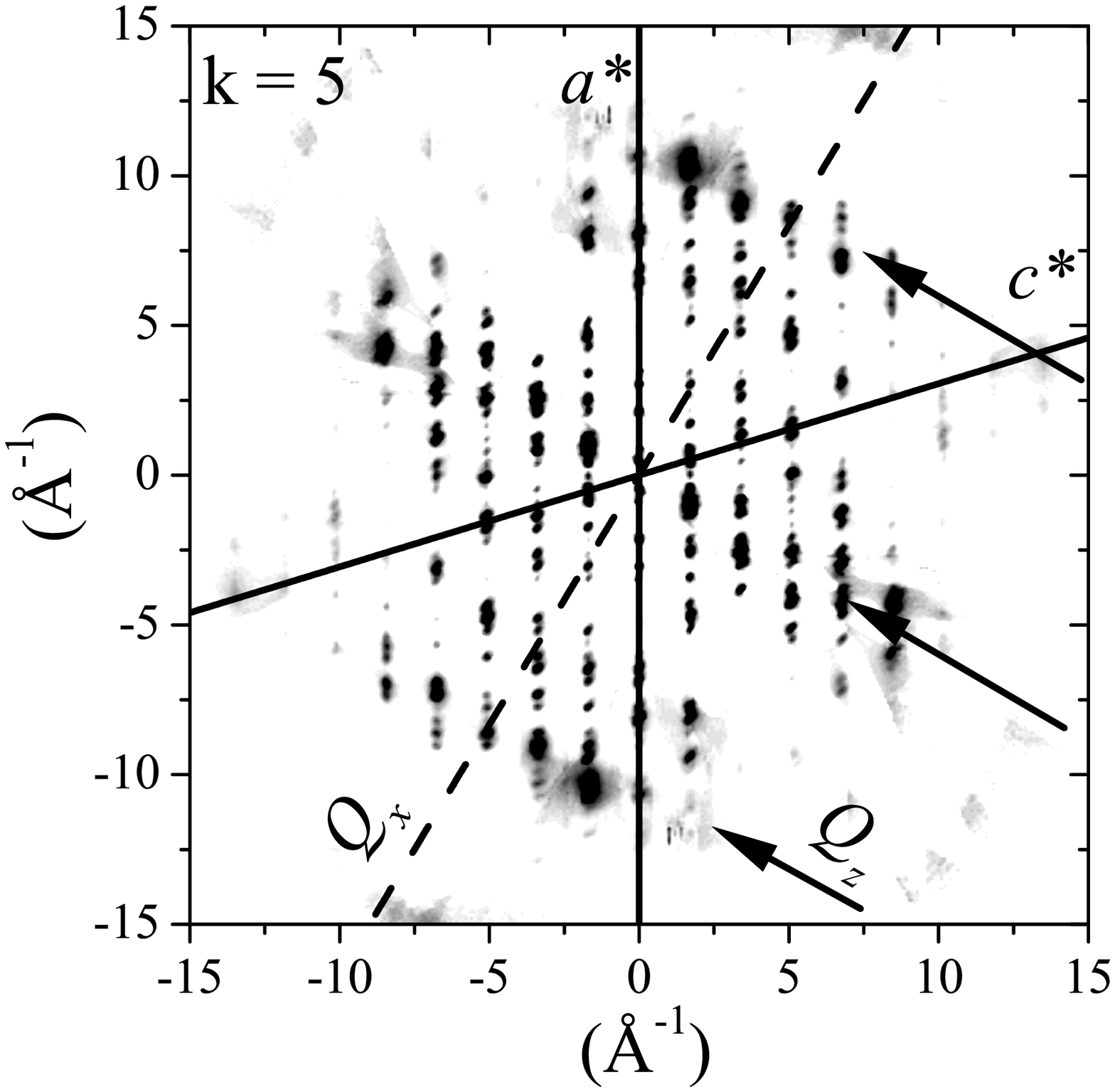}
\includegraphics[angle=0.,scale=0.40]{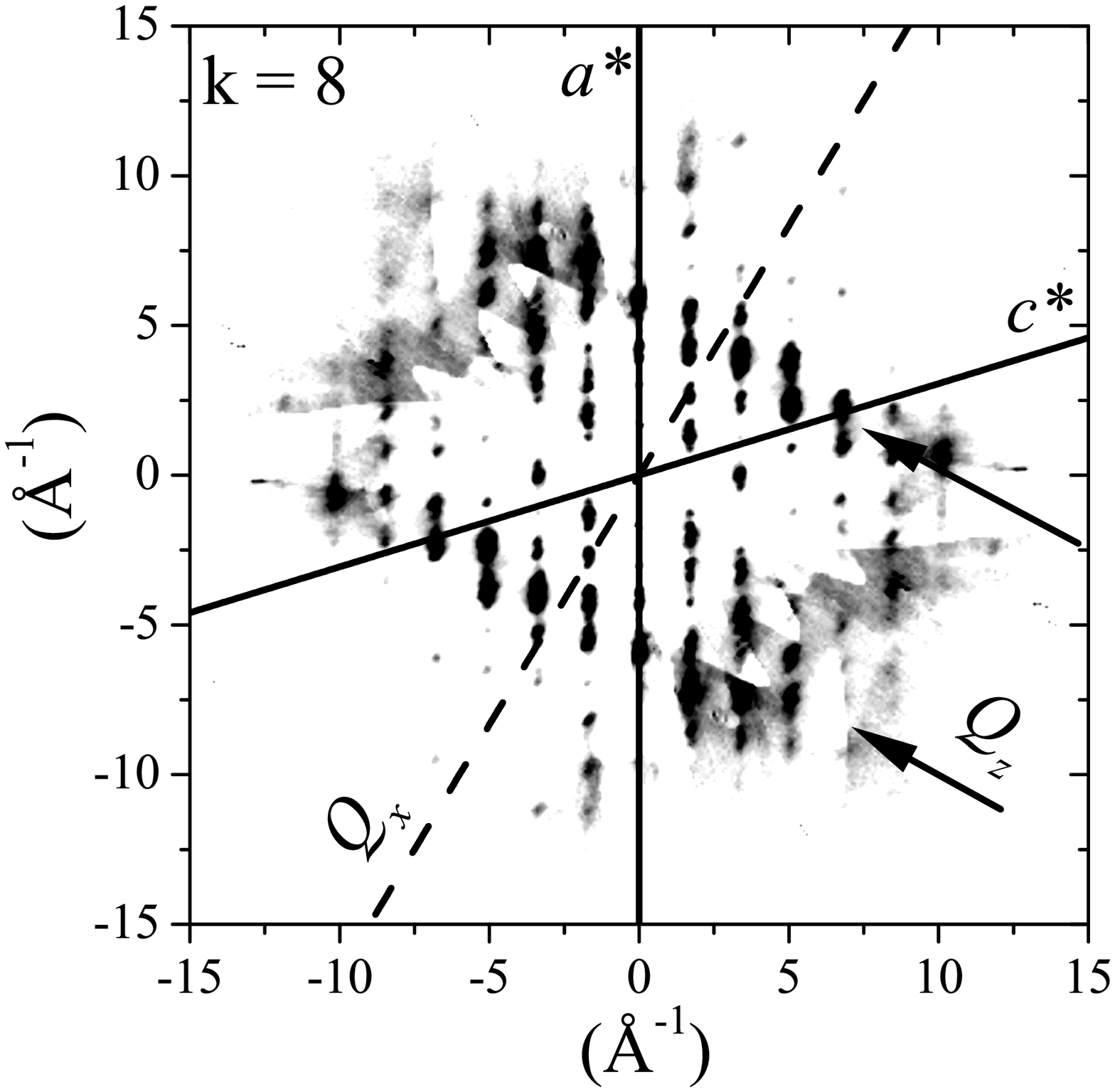}
\end{center}\caption{\label{fig:2} Diffraction patterns of KH$_{0.76}$D$_{0.24}$CO$_3$ at 300 K in ($a^*, c^*$) planes. The arrows point to the ridges of intensity for momentum transfer parallel to $(z)$ and perpendicular to the trace of dimer planes (dashed lines), which contain the $(x)$ direction, as defined in Fig. \ref{fig:1}. The $(y)$ direction parallel to ($b^*$) is perpendicular to $(a^*,c^*)$ planes. $k = Q_y/b^*$.} 
\end{figure}

The interpretation of the scattering function is based upon the crystal structure made of centrosymmetric dimers lying in $(103)$ planes with O$\cdots$O bonds parallel to each other. Proton/deuteron coordinates are $x,y,z,$ (Fig. \ref{fig:1}), $Q_x,Q_y,Q_z,$ are projections of $\Q$, $U_{xx}, U_{yy}, U_{zz}$, are the dominant temperature factors. Dotted lines in Fig. \ref{fig:1} enhance the network of double lines of proton/deuteron sites, either parallel to $(b)$, separated by $\Delta_x \approx 0.6$ \AA, or parallel to $x$, with $\Delta_y \approx 2.2$ \AA. The spatial periodicity of these grating-like structures is $D_x \approx a/\cos 42^\circ \approx 20.39$ \AA\ or $D_y = b$. The distance between $(x,y)$ planes is $D_z \approx 3.28$ \AA. A previous analysis of rigid body motions has shown that there is no short-range correlation between protons and carbonate entities \cite{FCKeen} and inelastic neutron scattering (INS) spectra show that proton dynamics are separated from heavy atoms and virtually dispersion-free \cite{FTP,IKSYBF}. 

In Fig. \ref{fig:2}, cuts of the scattering function parallel to $(a^*,c^*)$ show spots of intensity, corresponding to Bragg peaks, and, at large $\Q$-values, enhanced streaks along $Q_z$ perpendicular to the trace of dimer planes parallel to ($Q_x,Q_y$), with $Q_y || (b^*)$. These streaks are not an artefact of the instrument because: (i) the detectors were calibrated with vanadium; (ii) the streaks are measured by different detectors for each of the 8 orientations of the crystal; (iii) spurious signals eventually due to an anomalous detector should appear along different trajectories. 

The maps in Fig. \ref{fig:2} are similar to those of KHCO$_3$ at the same temperature \cite{FCG2}, apart from the weaker intensity of diffuse scattering due to the smaller concentration of protons. On the other hand, they are markedly different from those of KDCO$_3$ for which enhanced features are totally extinguished \cite{FCKeen}. The enhanced diffraction peaks correspond to nodes of the reciprocal sublattice of protons/deuterons and there is no obvious alternative sublattice with similar parameters. They occur when $k = Q_y/b^* = \pm 2\pi n_y /(b^*\Delta_y) \approx \pm n_y \times 2.57$. For $k = 0$, in-phase scattering by double lines of sites parallel to $x$ gives peaks at $Q_{x} = \pm 2\pi/\Delta_x = \pm (10.00 \pm 0.25)$ \A\ and $Q_z \approx n_z \times 1.92$ \A. At $k = \pm 3$, $n_y \approx 1$, anti-phase scattering along $Q_y$ and $Q_x$ give peaks at $Q_{x} = \pm (5.00 \pm 0.25)$ \A, $Q_z \approx n_z \times 1.92$ \A. For $k = \pm 5$ ($n_y \approx 2$) and $k = \pm 8$ ($n_y \approx 3$), the patterns of enhanced peaks are similar to those at $k = 0$ and $k = \pm 3$, respectively. 

\begin{figure}
\begin{center}\includegraphics[scale=0.6]{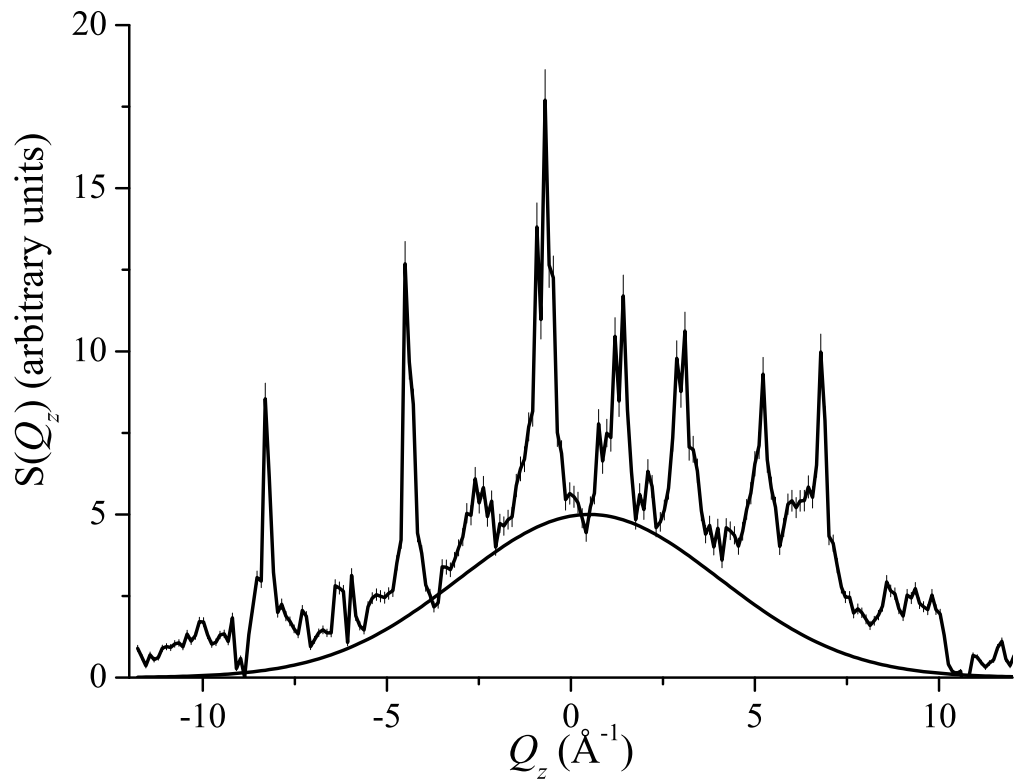}\end{center}
\caption{\label{fig:3} Cut of the scattering function along $Q_z$ in the plane $k = 0$, at $Q_x = (10 \pm 0.25)$ \A, after background correction. The Gaussian profile $\exp(-Q_z^2u_z^2)$, with $u_z^2 = 0.04$ \AA$^2$ is representative of the Debye-Waller factor for protons determined by neutron diffraction in Ref. \cite{FCou2}.}
\end{figure}

A cut along the streak at $Q_x = (10.00 \pm 0.25)$ \A\ in the plane $k = 0$ (Fig. \ref{fig:3}) shows enhanced Bragg peaks on the top of a continuum along $Q_z$. Because there is no evidence of disorder along $z$, this continuum is essentially due to incoherent scattering \cite{FCKeen}. It compares favorably with the Gaussian profile $I_{10}\exp-Q_z^2U_{zz}$, with $U_{zz} \approx 0.04$ \AA$^2$ \cite{FCou2}. Compared to the intensity $I_0$ at $Q_x = 0,\ Q_z =0$, the ratio $I_0/ I_{10} \approx 2$, with substantial uncertainty, is much greater than expected from conventional models according to which incoherent scattering by protons should be proportional to the Debye-Waller factor DWF $=\exp-(Q_x^2U_{xx} + Q_y^2U_{yy} + Q_z^2U_{zz})$, while incoherent scattering by other nuclei should be negligible. With $U_{xx} \approx 0.035$ \AA$^2$ \cite{FCou2}, the expected intensity ratio should be $I_0/ I'_{10} \approx 30$. The observed intensity ($I_{10}$) suggests that $U_{xx}$ is dramatically reduced for these $Q_x,Q_y$-values. 

\begin{figure}\begin{center}
\includegraphics[scale=0.5]{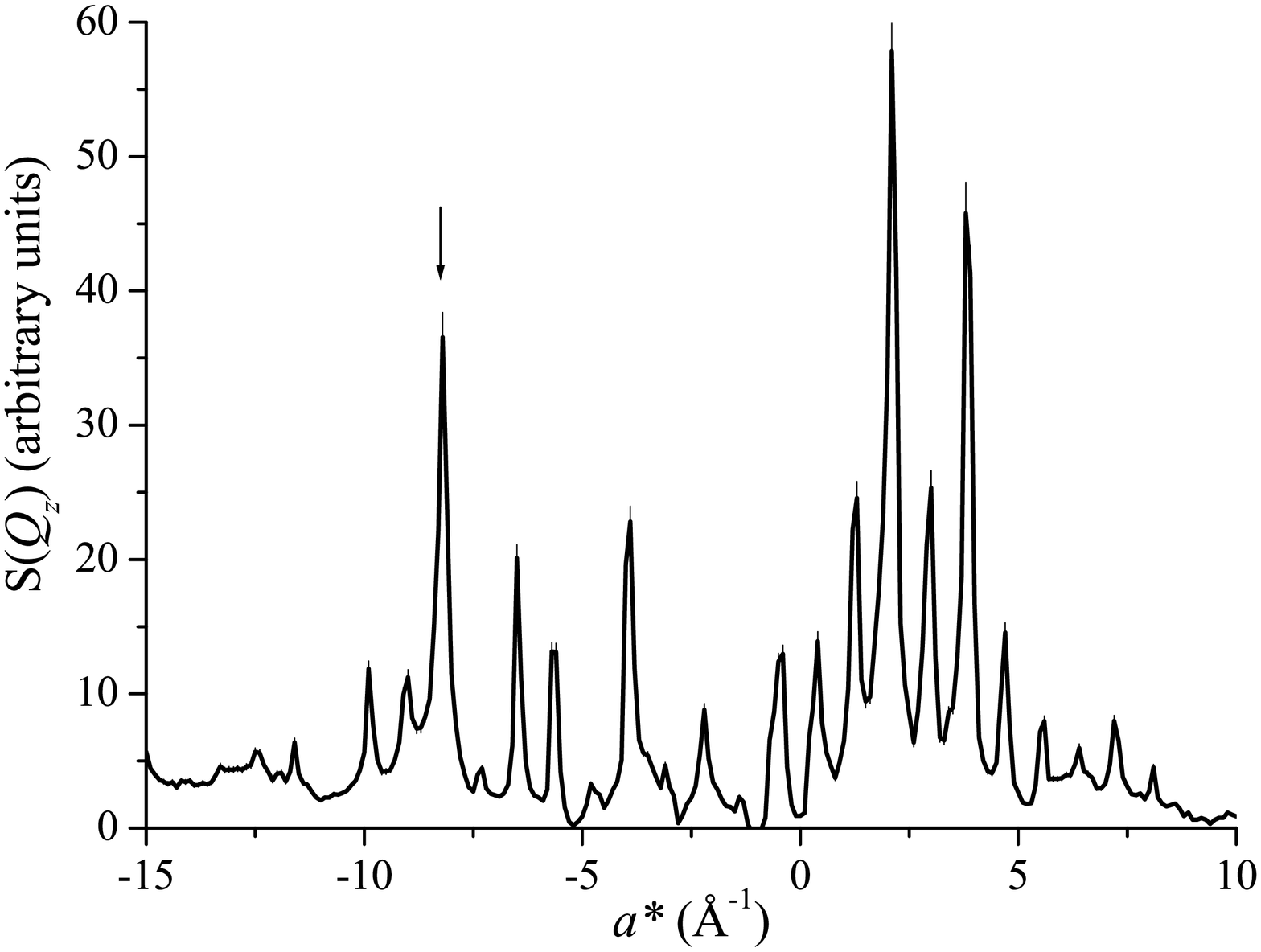}
\includegraphics[scale=0.5]{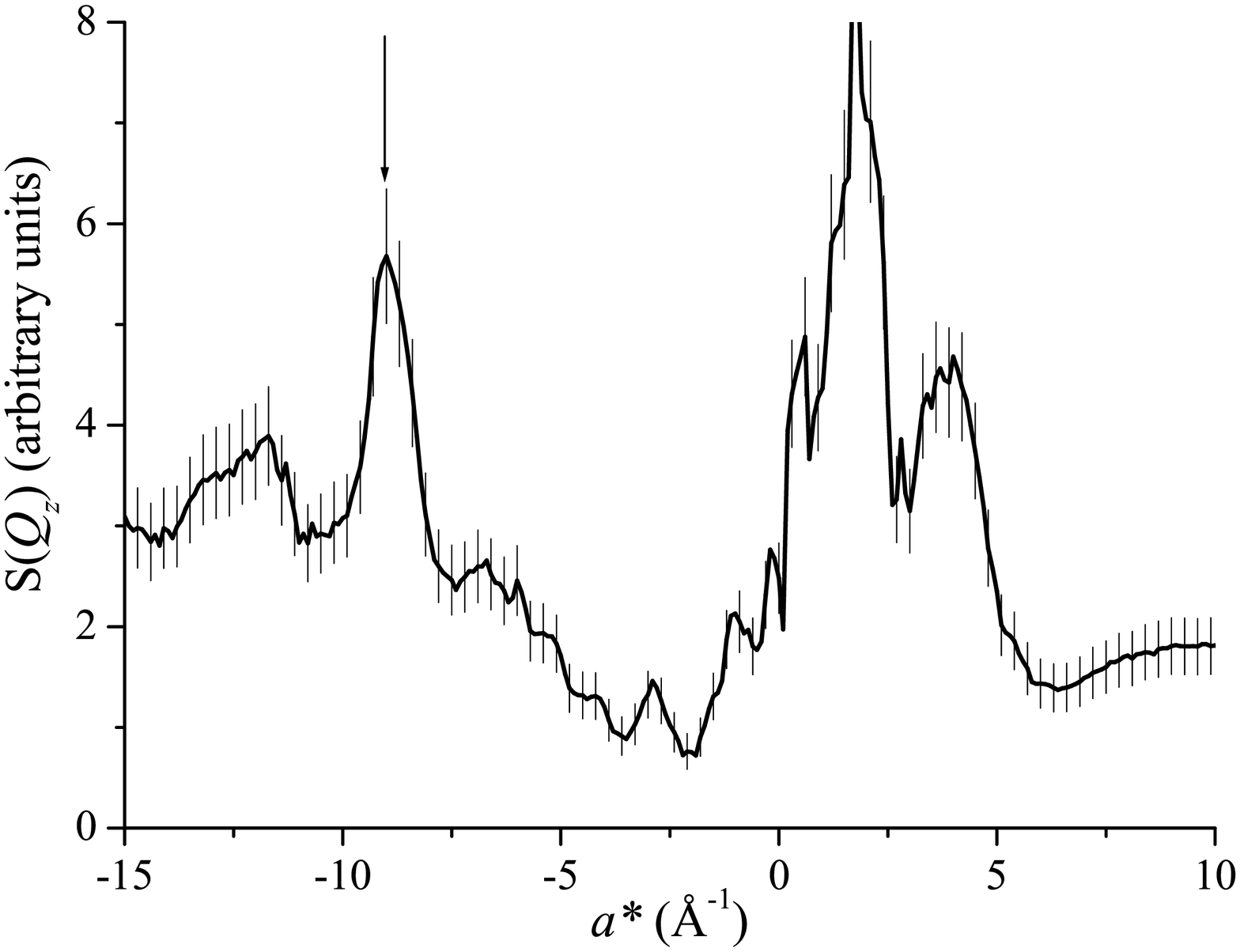}
\end{center}
\caption{\label{fig:4} Cuts of the scattering function along $a^*$ in the plane $k = 0$, at $-(5.0 \pm 0.2)$ \A (top) and $-(4.5 \pm 0.2)$ \A (bottom) off-$a^*$, after background correction. The arrows point to the enhanced Bragg peak (top) and to the diffuse scattering (bottom). }
\end{figure}

Fig. \ref{fig:4} (top) shows a cut at $k = 0$, parallel to $a^*$, through the reciprocal plane at $-(5.0 \pm 0.2)$ \A\ off-$a^*$. The enhanced Bragg peak observed at $-(8.5 \pm 0.2)$ \A\ (spotted by an arrow) hides the underlying incoherent scattering signal. Alternatively, a cut at $-(4.2 \pm 0.2)$ \A\ off-$a^*$ is free of Bragg peaks (Fig. \ref{fig:4}, bottom). It reveals the sharp profile of incoherent scattering centered at $-(9.0 \pm 0.2)$ \A\ (spotted by an arrow). Compared to the enhanced Bragg peak, the shift in position by $\approx -0.5$ \A\ is consistent with the slope of the $Q_z$ axis. The half-width at half-maximum, $\Delta_{a*} \approx 0.4 $ \A, gives the half-width along $Q_x$, $\Delta_{Qx} \approx \Delta_{a*} / \cos 42^\circ \approx 0.3 $ \A, similar to that of the Bragg peak ($\approx 0.2$ \A). Consequently, $U_{xx}$ vanishes exclusively if $Q_x$ matches a node of the reciprocal sublattice of protons (see below). Otherwise, incoherent scattering is not observed, presumably because it is dramatically depressed as $\exp-(Q_x^2U_{xx})$. Similarly, cuts around $k = 0$ (not shown) suggest that $U_{yy}$ vanishes sharply outside the plane. 

\subsection{Separation of proton and deuteron states: Raman scattering}

\begin{figure}
\begin{center}
\includegraphics[scale=0.6]{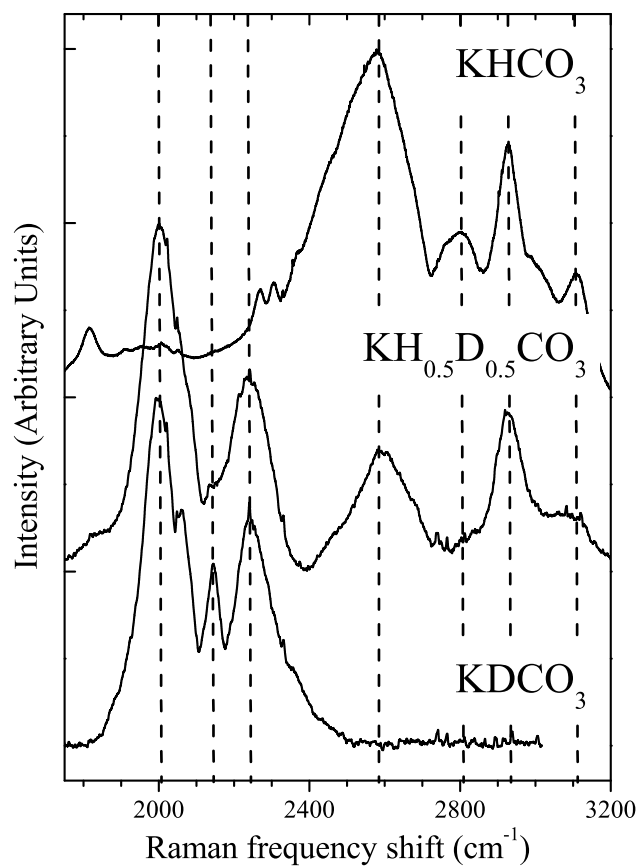}
\end{center}
\caption{\label{fig:5} Raman spectra of powdered crystal of KHCO$_3$ (top), KH$_{0.5}$D$_{0.5}$CO$_3$ (middle), and KDCO$_3$ (bottom) at 300 K. The dashed lines are guides for the eyes. Spectra were recorded with a triple-monochromator DILOR XY$\circledR$  equipped with an Ar$^+$ laser emitting at 4880 \AA. The spectral resolution was $\approx 2$ \cm. The samples were sealed in glass tubes. }
\end{figure}

In Fig. \ref{fig:5}, the Raman profile of the stretching mode $\nu$OD for the pure crystal is red-shifted by $\approx 600$ \cm\ with respect to the $\nu$OH profile of KHCO$_3$ and the spectrum of the isotope mixture is a mere superposition of these profiles. There is no visible frequency shift for each profile and, therefore, there is no significant dynamical correlation between isotopes. By contrast, if protons and deuterons were paired up, a single $\nu$O(H$_p$D$_{1-p}$) profile should appear at an intermediate frequency, shifting smoothly from $\nu$OH ($p = 1)$ to $\nu$OD ($p = 0$). This is clearly excluded. The spectra accord with separable proton and deuteron extended states consistent with the mixture of sublattices (\ref{eq:1}). 

In addition, the O$\cdots$O bond lengths are significantly different for KHCO$_3$ ($R_H = 2.58$ \AA) or KDCO$_3$ ($R_D = 2.61$ \AA) and $R_{0.76} = 2.59$ \AA\ for KH$_{0.76}$D$_{0.24}$CO$_3$ is seemingly a weighted average \cite{FCou2}. Isomorphous sublattices of protons and deuterons exclude a statistical distribution of O$\cdots$O distances. Otherwise, the $\nu$OH frequency should vary significantly with $p$, for $\Delta\nu \mathrm{OH} /\Delta R_p \approx 5000$ \cm\AA$^{-1}$ \cite{Novak}. The absence of frequency shift, or band broadening, suggests that a mixed crystal can be represented by isomorphous sublattices $p$[KHCO$_3$] and $(1-p)$[KDCO$_3$], whose structures and dynamics are independent of $p$. These sublattices are not resolved by diffraction. 

\section{\label{sec:3}Theory}

Admittedly the enhanced features are an evidence of quantum correlations, but different mechanisms have been envisaged: symmetry related entanglement \cite{FCKeen,FCG2,FCG4,FCG5}; quantum exchange \cite{KL3}; spin-spin interaction \cite{SOY}. They will be compared. 

The Raman spectra suggest that the Hamiltonian for a mixed crystal can be partitioned as 
\begin{equation}\label{eq:3}\begin{array}{rcl}
\mathcal{H}_p & = & p(\mathcal{H}_{\mathrm{H}} + \mathcal{H}_{\mathrm{KCO_{3H}}} + \mathcal{C}_{\mathrm{H,KCO_{3H}}}) \\
& + & (1-p)(\mathcal{H}_{\mathrm{D}} +\mathcal{H}_{\mathrm{KCO_{3D}}} + \mathcal{C}_{\mathrm{D,KCO_{3D}}}) + \mathcal{H}_{ep}.
\end{array}\end{equation}
$\mathcal{H}_{\mathrm{H}}$, $\mathcal{H}_{\mathrm{D}}$, represent the sublattices of H$^+$, D$^+$; $\mathcal{H}_{\mathrm{KCO_{3H}}}$, $\mathcal{H}_{\mathrm{KCO_{3D}}}$, those of heavy nuclei; $\mathcal{C}_{\mathrm{H,KCO_{3H}}}$, $\mathcal{C}_{\mathrm{D,KCO_{3D}}}$, couple light and heavy nuclei. The sublattices are bound through the electronic term $\mathcal{H}_{ep}$. There is no dynamical coupling between the KHCO$_3$ and KDCO$_3$ subsystems which are supposed to be separable: any state vector of $\mathcal{H}_p$ is a tensor product of state vectors of the subsystems. Furthermore, the adiabatic separation of $\mathcal{H}_{\mathrm{H}}$ and $\mathcal{H}_{\mathrm{KCO_{3H}}}$, or $\mathcal{H}_{\mathrm{D}}$ and $\mathcal{H}_{\mathrm{KCO_{3D}}}$, leads to factorable state vectors for the subsystem of bare protons (fermions), or deuterons (bosons) \cite{KL3,SOY,HMOY}. The separation of proton and deuteron states is, therefore, consistent with the ``super selection rule'' for fermions and bosons \cite{WWW}.

The mixed crystal is described with $N_a$, $N_{b}$, $N_{c}$ ($\mathcal{N}=N_{a}N_{b}N_{c}$) unit cells labelled $j,k,l,$ along crystal axes $(a),$ $(b),$ $(c)$, respectively. The sublattices are described with the same unit cell containing $4p$ KHCO$_3$ and $4(1-p)$ KDCO$_3$ entities. Proton/deuteron sites are organized in two dimer entities indexed as $j,k,l$ and $j',k,l$, respectively, with $j = j'$. For each dimer, dynamics are represented by normal coordinates, either antisymmetric $\{\alpha_{ar}\}$ or symmetric $\{\alpha_{sr}\}$ ($r = 1,2,3$), which are linear combinations of displacements along $x,y,z$, respectively. The coordinates of the centers of symmetry, $\{\alpha_{0r}\}$, $\{\alpha'_{0r}\}$, are $(x_0, y_0, z_0)$, $(x'_0, y'_0, z'_0)$. For planar dimers, $z_0 = z'_0 = 0$. Configurations $LL$ and $RR$ correspond to opposite signs for $x_0$ and $x'_0$. Although proton and deuteron sites may be not strictly identical, their dynamics are formally represented by the same set of normal coordinates. The wave functions in the ground state are represented by gaussian functions $\Psi_{0{jkl}}^{a} (\alpha_{ar})$ and $\Psi_{0{jkl}C}^{s}(\alpha_{sr} -\sqrt{2}\alpha_{0rC})$, with $C =$ ``$LL$'' or ``$RR$''. 

A key issue of the theory is to clearly differentiate sites and states. Sites refer to local positional parameters, whereas each single state is extended over all indistinguishable sites. Conversely, the probability density (occupancy) at any site comprises contributions from every single state. The nonlocal observables (normal coordinates and conjugated momentums) are not bound to any particular unit cell. We should emphasize that nonlocality due to the lattice periodicity is not related to any physical interaction. It is energy free. This is totally different in nature from the long-range quantum exchange in a supersolid, like the crystal of He \cite{BC1}. Obviousy, the proposal by Keen and Lovesey \cite{KL3} that nonlocality could arise from non vanishing exchange integrals is incorrect, even for the nearest sites separated by $\approx 2.2$ \AA\ \cite{FCou}. 

\subsection{\label{sec:31}Macroscopic states}

It is customary to represent lattice dynamics with plane waves (phonons). For deuterons in the ground state, this can be written as: 
\begin{equation}\label{eq:4}
|\mathrm{D}^+_{\mathbf{k}\varsigma}\rangle_C = \displaystyle{\frac{1} {\sqrt {\mathcal{N}}}}  \sum\limits_{{l} = 1}^{{N_c}} \sum\limits_{{k} = 1}^{{N_b}} \sum\limits_{{j} = 1}^{{N_a}} |\Xi_{0Djkl\varsigma C} \rangle \exp(i\mathbf{k\cdot L}_{jkl}); 
\end{equation}
where $\mathbf{k}$ is the wave vector; $\mathbf{L}_{jkl}  = j \mathbf{a} + k \mathbf{b} + l \mathbf{c}$ is a lattice vector; $\mathbf{a}$, $\mathbf{b}$, $\mathbf{c}$, are the unit cell vectors. The wave functions for a unit cell, $\Xi_{0Djkl\varsigma C} = \Theta_{0D{jkl}C} +\varsigma \Theta_{0D{j'kl}C}$ with $\varsigma =$ ``$+$'' or ``$-$'', is a linear combination of the wave functions for centrosymmetric sites: $\Theta_{0DjklC} (\{\alpha_{0r}\}) = \prod\limits_r \Psi_{0D{jkl}}^{a} (\alpha_{ar}) \Psi_{0D{jkl}C}^{s}(\alpha_{sr} -\sqrt{2}\alpha_{0rC})$ and likewise for $\Theta_{0D{j'klC}}(\{\alpha'_{0r}\}) $. For centrosymmetric dimers, there is no permanent dipole, so long-range correlations through inter-dimer interaction are negligible. Phonons are virtually dispersion-free \cite{FTP,IKSYBF}. 

$|\mathrm{D}^+_{\mathbf{k}\varsigma}\rangle_C$ is a single-deuteron state with an occupancy number $(4\mathcal{N})^{-1}$ per site. This state cannot be factored with eigen states for local coordinates ($x_{1jkl}, x_{2jkl} \ldots $). It is entangled in position. (By definition, a pure state of an arbitrary number of particles is said to be entangled when, because of its indivisible nature, it cannot be described by treating each particle separately.) Likewise, the state in momentum space is also entangled. In addition, this state meets the symmetry postulate for bosons stipulating that the sign should not change when any two dimer sites, say $j,k,l,$ and $j'',k'',l''$, or any two sites within dimers, are interchanged \cite{CTDL}. 

Alternatively, eigen states for fermions should be antisymmetric. An interchange of lattice sites $j,k,l,$ and $j'',k'',l'',$ should transform a plane wave into the anti-phase wave and superposition of these waves should annihilate the amplitude. Hence, phonons are forbidden, except those matching the lattice periodicity, for which any interchange is neutral. Such states represent in-phase dynamics in each unit cell, namely $\varsigma =$ ``$+$'', and in-phase dynamics for all unit cells, namely $\mathbf{k\cdot L}_\mathrm{H} \equiv 0 \mathrm{\ modulo\ } 2\pi$, where $\mathbf{L}_\mathrm{H}$ is any vector of the sublattice of protons. This macroscopic quantum effect excluding elastic distortions has been named ``super-rigidity'' \cite{FCG2}. In addition, eigen states must be antisymmetrized with respect to interchange of proton sites within dimers, what is equivalent to changing the signs of the coordinates for the center of symmetry, say $\{\alpha_{0r}\} \longrightarrow -\{\alpha_{0r}\}$. To this end, the wave function of a dimer, supposed to be non planar for the ease of notation, is rewritten as a linear combination of the wave functions for permuted sites as: $\Theta'_{0H{jkl}C\tau} = 2^{-1/2}[\Theta_{0HjklC} (\{\alpha_{0r}\}) + \tau \Theta_{0HjklC} (-\{\alpha_{0r}\})]$ (and likewise for $j'$), with $\tau =$ ``$+$'' or ``$-$''. Then, antisymmetrization requires a spin-symmetry, either singlet-like ($S$ for $\tau=+$) or triplet-like ($T$ for $\tau=-$). Finally, with the translation invariant wave function of the unit cell, $\Xi'_{0H{jkl}+C\tau} = \Theta'_{0H{jkl}C\tau} +\Theta'_{0H{j'kl}C\tau}$, protons states can be written as:
\begin{equation}\label{eq:5}
\begin{array}{c}
|\mathrm{H}^+_+\rangle_C = \displaystyle{\frac{1}{\sqrt {\mathcal{N}}}}  \sum\limits_{{l} = 1}^{{N_c}} \sum\limits_{{k} = 1}^{{N_b}} \sum\limits_{{j} = 1}^{{N_a}} |\Xi'_{0H{jkl}+C+} \rangle |S \rangle; \\
|\mathrm{H}^+_-\rangle_C = \displaystyle{\frac{1}{\sqrt {\mathcal{N}}}}  \sum\limits_{{l} = 1}^{{N_c}} \sum\limits_{{k} = 1}^{{N_b}} \sum\limits_{{j} = 1}^{{N_a}} |\Xi'_{0H{jkl}+C-} \rangle |T \rangle.
\end{array}
\end{equation}
For each configuration $C$, the ground states are entangled in position, momentum, and spin. The singlet-like and triplet-like states are degenerate because the spin symmetry is energy-free. The proposal by Sugimoto et al. \cite{SOY} that the spin symmetry could be due to spin-spin interaction is incorrect because the actual coupling between nearest spin sites ($\sim 10^4$ Hz $\sim 10^{-6}$ K) is far too weak to give rise to long-lived correlations. 

For an isolated crystal, symmetry-related quantum-correlations are decoherence-free. Thanks to the adiabatic separation, there is no phonon-induced decoherence and, therefore, no transition to the classical regime, before melting or decomposition. In practice, some states can be transitorily disentangled by external perturbations, such as thermal radiations, but this is counterbalanced by spontaneous re-entanglement via decay to the ground state. In a standard gaseous atmosphere, the density of the surroundings is $\sim 1000$-fold smaller than that of the crystal, so the impact of environment-induced decoherence to measurements is invisible. A source of disentanglement is the thermal population of excited proton states (for example $\gamma_a$ OH and $\gamma_s$ OH at $\approx 1000$ \cm), as the $a-s$ splitting destroys the spin-symmetry, but this is marginal. 

Because the spin symmetry holds for both $LL$ and $RR$ ground states, for which there is no $a-s$ splitting, the sublattice at 300 K can be represented by a mixture comprising a coherent superposition of super-rigid states for protons (\ref{eq:5}), on the one hand, and, on the other, a coherent superposition of phonon states for deuterons (\ref{eq:4}):
\begin{equation}\label{eq:6}
\begin{array}{l}
|H^+(T)\rangle = \sum\limits_\tau [q'_{LLH}(T) |\mathrm{H}^+_{\tau}\rangle_{LL} + q'_{RRH}(T) |\mathrm{H}^+_{\tau}\rangle_{RR}]; \\
|D^+_\mathbf{k}(T)\rangle = \sum\limits_\varsigma [q'_{LLD}(T)|\mathrm{D}^+_{\mathbf{k}\varsigma }\rangle_{LL} + q'_{RRD}(T) |\mathrm{D}^+_{\mathbf{k}\varsigma }\rangle_{RR}]; 
\end{array}\end{equation}
with $|q'_{LLH}(T)|^2 = q_{LLH}(T)$, etc, in accordance with (\ref{eq:1}). 

\subsection{\label{sec:32}Neutron scattering}

The partial differential nuclear cross-section can be written as \cite{SWL} 
\begin{equation}\label{eq:7}
\frac{\mathrm{d}^2\sigma}{\mathrm{d}\Omega\mathrm{d}E} = \frac{k_f}{k_i} \sum\limits_{i,f} p_i |\langle\mathbf{k}_f|\langle f|\hat{V}(\mathbf{r})|i\rangle|\mathbf{k}_i\rangle|^2 \delta(\hbar\omega + E_i -E_f) .
\end{equation}
The summation runs over initial states $|i\rangle$, $E_i$, weight $p_i$, and final states $|f\rangle$, $E_f$. The energy transfer is $\hbar \omega$ and $\mathrm{d}\Omega$ is an element of solid angle. The interaction potential between the incident neutron and the target sample, $\hat{V}(\mathbf{r})$, comprises the interaction potential at every site. For diffraction, $|i\rangle \equiv |f\rangle$, $E_i \equiv E_f$, $\hbar\omega \equiv 0$. Neutrons are diffracted by each single state. Because of the adiabatic separation, neutrons diffracted by proton or deuteron states are free of any contribution from heavy atoms. Moreover, neutrons diffracted by proton states probe exclusively the interaction potential at the proton sites and likewise for deuteron states, irrespective of whether or not these sites are identical. There is no Laue monotonic diffuse scattering. In general, $\hat{V}(\mathbf{r})$ does not have the periodicity of the lattice, for the potential will depend upon the nuclear position and spin orientation. However, this is not the case for the super-rigid states which match exactly the periodicity of the crystal. 

Upon momentum transfer to the target proton states (\ref{eq:6}), the final state is, in general, such that $\mathbf{k}_f\mathbf{\cdot L}_\mathrm{H} \neq 0 \mathrm{\ modulo\ } 2\pi$, so elastic distortions are generated. They are represented by ellipsoids at H/D sites in Fig. \ref{fig:1}. The super-rigid states are probed without induced distortions when 
\begin{equation}\label{eq:8}
\mathbf{Q\cdot L}_\mathrm{H} = 0 \mathrm{\ modulo\ } 2\pi. 
\end{equation}
Then, neutrons probe the spin-symmetry and the coherent cross-section is the total nuclear cross-section, $\sigma_H \approx 82.0$ b, that is $\approx 45$ times greater than the coherent cross-section for regular Bragg peaks, $\sigma_{Hc} \approx 1.8$ b, or $\approx 15$ times greater than $\sigma_{Dc} \approx 5.59$ b, or $\approx 3$-fold greater than $\sigma_{\mathrm{KCO_3}c} \approx 27.7$ b for heavy atoms. These scattering events emerge from the diffraction pattern with spectacular contrast, even at large $|\Q|$-values as they are not attenuated by the Debye-Waller factor. Besides, they are temperature independent. 

For the $P2_1/a$ structure, $z$ is not parallel to an axis of the sublattice of protons (Fig. \ref{fig:1}), so the matching condition (\ref{eq:8}) is never realized along $Q_z$. The super-rigidity along $z$ is necessarily destroyed by measurements and intensities are proportional to $\exp-U_{zz}Q_z^2$. On the other hand, since $z_0 =z'_0 =0$ for planar dimers, the spin-symmetry holds only in the $(x,y)$ planes, so the differential cross-section for enhanced diffraction can be written as \cite{FCG4} 
\begin{equation}\label{eq:9}
\begin{array}{rcl}
\displaystyle{\frac{d\sigma_c}{d\Omega}} & \propto & \sum\limits_{\tau_{i}} \sum\limits_{\tau_{f}} \left | \sum\limits_{j = 1} ^{N'_a} \sum\limits_{k =1} ^{N_b} \sum\limits_{l =1} ^{N_c} \exp i Q_z l D_z \right. \left\{\left [ \exp iQ_y \left(kD_y - \Delta_y/2 \right)\right.\right.\\
& + & \left. \tau_{i}\tau_{f} \exp iQ_y \left(kD_y + \Delta_y/2 \right ) \right ] \\
& \times & \left. \left [\exp i Q_x \left(jD_x - \Delta_x/2 \right ) + \tau_{i}\tau_{f} \exp i Q_x \left(jD_x + \Delta_x/2 \right ) \right ] \right\}^2 \Bigg |^2 \\
& \times & \exp(-U_{zz}Q_z^2) \times \delta(\omega), 
\end{array} 
\end{equation}
where $\{\cdots\}^2$ takes into account indistinguishable scattering events, either in-phase ($|S\rangle \longrightarrow |S\rangle$ and $|T\rangle \longrightarrow |T\rangle$), or anti-phase ($|S\rangle \longrightarrow |T\rangle$ and $|T\rangle \longrightarrow |S\rangle$).  Enhanced diffraction by $(x,y)$ planes is anticipated at $Q_{Hy} = \pm 2\pi n_y/\Delta_y$, $Q_{Hx} = \pm 2\pi n_x/\Delta_x$ for $n_y$ even (in-phase scattering) and $Q_{Hx} = \pm 2\pi (2n_x+1)/\Delta_x$, for $n_y$ odd (anti-phase scattering). Then, enhanced Bragg peaks occur at $Q_z = 2\pi n_z/D_z$, with intensities proportional to $\exp-U_{zz}Q_z^2$. 

Otherwise, if $Q_z \neq 2\pi n_z/D_z$, the super-rigid $(x,y)$ lattice is probed without distortion for the same $(Q_{Hx},Q_{Hy})$-values, but the spin-symmetry is destroyed. Incoherent scattering is observed and the intensity is proportional to 
\begin{equation}\label{eq:10}
\sigma_{Hi}\exp -(Q_z^2 U_{zz})\delta(Q_x - Q_{Hx}+ Q_y - Q_{Hy}), 
\end{equation}
hence the cigar-like shape. Compared to a nonrigid lattice, the enhancement is merely due to $U_{xx} = U_{yy} = 0$. The observed intensity ratio $I_0/I_{10} \approx 2$ is at variance with the expected ratio of $\approx 1.5$, but this discrepancy could be representative of errors in the background correction. 

Figures \ref{fig:2}-\ref{fig:4} can be thus rationalized with three different kinds of scattering events: (i) regular Bragg diffraction by the crystal lattice; (ii) enhanced diffraction by the fully entangled sublattice of protons; (iii) incoherent scattering by the super-rigid planes of protons. 

\section{\label{sec:4}Discussion}

The proposed theory is based upon fundamental laws of quantum mechanics applied to the crystal in question: the structure is periodic; dimers are centrosymmetric; indistinguishable protons are fermions; indistinguishable deuterons are bosons. It leads to macroscopically entangled states and, in the special case of protons, to super-rigidity and spin-symmetry with observable consequences. This theory is consistent with a large set of experimental data (neutron diffraction, QENS, INS, infrared and Raman) and, to the best of our knowledge, there is no conflict with any observation. There is, therefore, every reason to conclude that the crystal is a macroscopic quantum system for which only nonlocal observables are relevant. 

Because quantum entanglement emerges naturally from the crystal structure, there is no reason to introduce any further interaction. From the viewpoint of economy (Ockham's razor), alternative theories based upon quantum exchange \cite{KL3} or spin-spin interaction \cite{SOY} are less satisfactory. Moreover, these correlations are very unlikely to be significant and it is doubtful that these short-range interactions, essentially confined to dimers, could account for macroscopic correlations, especially at elevated temperatures. 

Protons are unique to demonstrating quantum entanglement, because they are fermions and because the very large incoherent cross-section can merge into the total coherent cross-section. No other nucleus can manifest such an increase of its coherent cross-section. The enhanced features can be, therefore, unambiguously assigned to protons, in accordance with their positions in reciprocal space. They are evidences of macroscopic quantum correlations which have no counterpart in classical physics. 

According to the theory, the pair correlation function for protons represented by ellipsoids in Fig. \ref{fig:1} are due to measurement induced phonons. The structure free of induced distortions should comprise Dirac functions for protons and ellipsoids for deuterons. The spatial wave functions for protons (\ref{eq:5}) describe the distribution of the center of mass of the proton states with respect to that for all other atoms and the uncertainty principle holds at each site of the perfectly correlated sublattice. 

It is widely believed that macroscopic objects are, in principle, governed by the laws of quantum mechanics and classical physics is an emergent phenomenon at an appropriate limit when the system interacts with an environment that might as well consist of its internal degrees of freedom \cite{Zurek}. The crystal of bicarbonate is a counterexample that is never amenable to classical physics, because (i) quantum entanglement is as robust as the crystal; (ii) the adiabatic separation insulates the proton states from every decoherence mechanism; (iii) the energy gap for excited proton states is much greater than $kT$. 

A different approach places the emergence of classicality on the impossibility of distinguishing individual quantum levels with coarse-grained measurements \cite{KB}. Transposed to diffraction, the interaction potential in (\ref{eq:7}) is replaced by the mean potential, $\bar{V}(\mathbf{r}) = \overline{\sum\limits_{i} p_i \langle i|\hat{V}(\mathbf{r})|i\rangle}$, that is then Fourier transformed into the unit-cell structure factor \cite{SWL}. Eigen states and nonlocal observables are thus translated into classical variables for particles located at definite sites and neutrons scattered by various nuclei can interfere, irrespective of whether or not these nuclei are involved in the same eigen states. Because of the opposite signs of the scattering lengths for protons and deuterons, interferences could build up in such a way that some Bragg peaks would be more intense for KHCO$_3$ than for KDCO$_3$, or vice versa. However, these enhanced peaks are not expected to single out the reciprocal sublattice of protons and the structure factor does not account for the enhanced incoherent scattering along $Q_z$ in any way. In addition, the rigid-body analysis does not provide any evidence of correlation between protons and heavy atoms \cite{FCKeen}. There is, therefore, experimental support to conclude that the assumption of classicality emerging from coarse-grained measurements is only a convenient approximation hiding the quantum nature of the crystal. Nevertheless, quantum effects embedded in the Bragg peaks transpire as the mean scattering length (\ref{eq:2}) is less satisfactory than the mixture of distinct sublattices (\ref{eq:1}), although the difference is not large enough to be firmly conclusive. 

\section{Conclusion} 

Neutron diffraction and Raman spectra show that the mixed isotope crystal is not a statistical mixture of KHCO$_3$ and KDCO$_3$ entities. The crystal is a macroscopic quantum system made of separable matter waves over the isomorphous sublattices of KHCO$_3$ and KDCO$_3$. The quantum theory shows that deuteron (boson) states can be represented by phonons whereas antisymmetrized proton (fermion) states demonstrate super-rigidity and spin-symmetry. This macroscopic quantum entanglement is energy free and decoherence free. It is as robust as the crystal structure and the emergence of the classical regime at elevated temperatures is forbidden. The main outcome of the present work is that macroscopic quantum effects for protons are not destroyed by deuteron states. Despite their different masses, isotopes are not defects leading to localization, possibly because they are isoelectronic. 

The quantum theory of diffraction by proton states allows us to distinguish different events. First, in the most general case, momentum transfer destroys the quantum entanglement and regular Bragg peaks are observed. Second, for those particular $\Q$-values matching the nodes of the reciprocal sublattice, the super-rigidity and the spin-symmetry are probed without any measurement induced symmetry breaking. The incoherent cross-section merges into the total coherent cross-section and the corresponding peaks show markedly enhanced intensities featuring macroscopic quantum entanglement. Because of the anomalous Debye-Waller factor, these peaks are observed at large $\Q$-values and they are temperature independent. Third, for those $\Q$-values matching only the nodes of the reciprocal planes of protons, the spin-symmetry is destroyed and cigar-like shaped rods of incoherent scattering by the super-rigid planes are observed in between the enhanced peaks. Clearly, the unit-cell structure factor is not appropriate to account for these enhanced features. 

This work presents one single case of macroscopically entangled states on the scale of Avogadro's constant in a mixed isotope crystal at room temperature. The quantum theory suggests that such macroscopic quantum effects should be of significance for many hydrogen bonded crystals. Hopefully, unforeseeable consequences of the quantum view advocated in this work will appear in the future.

\clearpage

\clearpage

\clearpage

\end{document}